\documentclass[prl, superscriptaddress,twocolumn, amsmath, amssymb, showpacs]{revtex4-2}
\usepackage{CJK}
\usepackage{color}

\usepackage{sidecap}
\usepackage{times}
\usepackage{verbatim}
\usepackage{graphicx}
\usepackage{color}
\usepackage{graphics}
\usepackage{amsmath}
\usepackage{float}

\usepackage{amssymb}
\usepackage[colorlinks,linkcolor=blue,
            citecolor=blue,
            hyperindex,
            pdfstartview=FitH,
            plainpages=false]
            {hyperref}

\begin{document}
\title{Unusual band splitting and superconducting gap evolution with sulfur substitution in FeSe}

\author{Yuanyuan Yang}
\affiliation{Key Laboratory of Artificial Structures and Quantum Control (Ministry of Education), Shenyang National Laboratory for Materials Science, School of Physics and Astronomy, Shanghai Jiao Tong University, Shanghai 200240, China}
\author{Qisi Wang}
\affiliation{State Key Laboratory of Surface Physics and Department of Physics, Fudan University, Shanghai 200433, China}
\author{Shaofeng Duan}
\affiliation{Key Laboratory of Artificial Structures and Quantum Control (Ministry of Education), Shenyang National Laboratory for Materials Science, School of Physics and Astronomy, Shanghai Jiao Tong University, Shanghai 200240, China}
\author{Hongliang Wo}
\affiliation{State Key Laboratory of Surface Physics and Department of Physics, Fudan University, Shanghai 200433, China}
\author{Chaozhi Huang}
\author{Shichong Wang}
\author{Lingxiao Gu}
\affiliation{Key Laboratory of Artificial Structures and Quantum Control (Ministry of Education), Shenyang National Laboratory for Materials Science, School of Physics and Astronomy, Shanghai Jiao Tong University, Shanghai 200240, China}

\author{Dong Qian}

\affiliation{Key Laboratory of Artificial Structures and Quantum Control (Ministry of Education), Shenyang National Laboratory for Materials Science, School of Physics and Astronomy, Shanghai Jiao Tong University, Shanghai 200240, China}
\author{Jun Zhao}
\affiliation{State Key Laboratory of Surface Physics and Department of Physics, Fudan University, Shanghai 200433, China}
\affiliation{Institute of Nanoelectronics and Quantum Computing, Fudan University, Shanghai 200433, China}
\author{Wentao Zhang}
\email{wentaozhang@sjtu.edu.cn}
\affiliation{Key Laboratory of Artificial Structures and Quantum Control (Ministry of Education), Shenyang National Laboratory for Materials Science, School of Physics and Astronomy, Shanghai Jiao Tong University, Shanghai 200240, China}

\date {\today}

\begin{abstract}

High-resolution angle-resolved photoemission measurements were taken on FeSe$_{1-x}$S$_x$ (x=0, 0.04, and 0.08) superconductors. With an ultrahigh energy resolution of 0.4 meV, unusual two hole bands near the Brillouin-zone center, which was possibly a result of additional symmetry breaking, were identified in all the sulfur-substituted samples. In addition, in both of the hole bands highly anisotropic superconducting gaps with resolution limited nodes were evidenced. We find that the larger superconducting gap on the outer hole band is reduced linearly to the nematic transition temperature while the gap on the inner hole is nearly S-substitution independent.
Our observations strongly suggest that the superconducting gap increases with enhanced nematicity although the superconducting transition temperature is not only governed by the pairing strength, demonstrating strong constraints on theories in the FeSe family.

\end{abstract}

\maketitle

In iron-based superconductors, the low energy excitation is governed by the five Fe 3$d$ orbits, resulting in complicated Fermi surface topologies in the Brillouin zone \cite{Lebegue2007}. The distribution of the superconducting gap on each Fermi surface sheet and the scattering of the Cooper pairs among different Fermi surface sheets are governed by the electron pairing symmetry \cite{Kreisel2020a}, asking for high-resolution momentum-resolved experimental technique to resolve them.
Electronic nematicity is common in high-temperature superconductors, and the role of the nematic phase played in high-temperature superconducting pairing is a predominant topic in studying the superconducting mechanism of the cuprate and pnictide  \cite{Fradkin2010}, the recently discovered Kagome-lattice-based superconductor \cite{Nie2022}, and the twisted bilayer superconductor \cite{RubioVerdu2022}.
Nematic order in the iron-based superconductors refers to an electronic phase with broken rotational symmetry but preserved translational symmetry\cite{Paglione2010,Chuang2010,Bohmer2015,Yi2011,Fu2012,Lu2014}.
Since there is universal nematic phase in the phase diagram of iron-based high-temperature superconductors and it may origin differently in different families, the interplay between the nematicity and the superconductivity is under hot debates in the field \cite{Hu2012,Si2016}.

Among the iron-based superconductors, FeSe holds the simplest crystal structure with the appearance of nematicity and also superconductivity. Moreover, the absence of the long-range magnetic order makes it an ideal system in studying the interplay between the nematicity and the superconducting pairing \cite{Hsu2008,Bohmer2015,McQueen2009,Bendele2010,Baek2015}.
In addition, phases in FeSe can be tuned effectively by applying physical or chemical pressure, intercalating or gating method, and growing monolayer FeSe on suitable substrates. 
Isoelectronic substitution, usually by replacing the Se with S, i.e., the FeSe$_{1-x}$S$_x$ system, is favored for tuning the phase diagram under normal conditions, and has been widely investigated by transport, photoemission, tunneling, and so on \cite{Coldea2019,AbdelHafiez2015,Wang2017,Watson2015a,Reiss2017,Xu2016,Tetsuo2021}. It is generally believed that a hole band near the Brillouin-zone center and an electron band near the Brillouin-zone boundary are responsible for the superconducting state, and a highly anisotropic energy gap is developed on the hole band although there is no consensus on a node or nodeless gap \cite{Kreisel2020a}. Recently, two hole pockets, which possibly results from an additional unknown symmetry breaking, are identified in FeSe \cite{Li2020a}. However, the two-hole electronic structure and the superconducting gap as a function of the S substitution are barely studied, possibly limited by experimental resolution due to the low superconducting transition temperature ($T_c\sim$10 K), small energy gap (2$\sim$3 meV), and the naturally twinned sample.

In this Letter, we report two hole pockets below $T_c$ near the Brillouin-zone center in FeSe$_{1-x}$S$_x$ ($x$ = 0, 0.04, and 0.08) and the superconducting gap distribution on each pocket by ultrahigh resolution angle-resolved photoemission spectroscopy (ARPES). In particular, with high energy resolution, we find the energy gap behaves differently on the two hole bands that the larger superconducting gap is reduced linearly to the nematic transition temperature while the other one is nearly S-substitution independent, although the superconducting transition temperature is higher with more S substitution. From the observations, we conclude that the appearance of nematicity favors to enhance the superconducting paring in FeSe$_{1-x}$S$_x$.
  
Ultrahigh resolution ARPES measurements were taken on a home-built 7 eV laser system cooperating with the time-resolved function \cite{Yang2019}. The overall energy resolution in the energy gap measurements was optimized to 0.4 meV with a beam size of about 30 $\mu m$ (full width at half maximum). High quality FeSe$_{1-x}$S$_x$ single crystals were grown using KCl-ACl$_3$ flux under a permanent temperature gradient \cite{Wang2016a}, and cleaved under ultrahigh-vacuum conditions with a pressure of better than 3$\times$10$^{-11}$ Torr. Thermal drift of the sample position was automatically corrected by a computer with a precision less than 1 $\mu m$, ensuring the temperature dependent measurements to be taken on a fixed spot of the cleaved surface. The beam spot was focused mostly on a single detwinned domain in the experiments.

\begin{figure}
\centering\includegraphics[width=1\columnwidth]{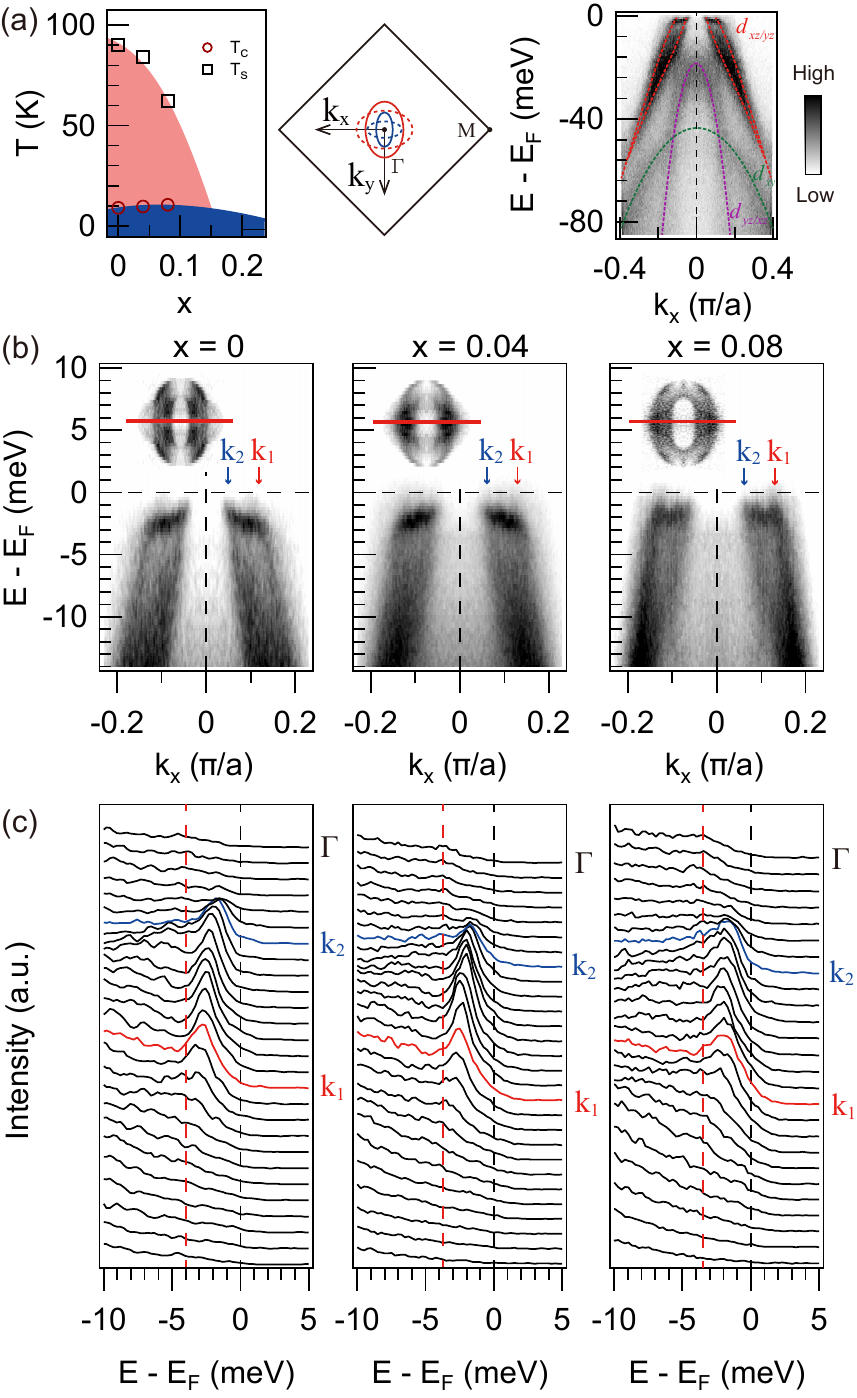}
\caption{
Photoemission data of FeSe$_{1-x}$S$_x$ ($x$ = 0, 0.04, 0.08) samples measured along $\Gamma$-M near the Brillouin zone center at 4.5 K.
(a) Phase diagram, Brillouin zone, and the $\Gamma$-M photoemission spectrum of FeSe. The $T_c$ and the $T_s$ are determined by transport measurements and temperature dependent photoemission experiments (Fig.~\ref{Fig3}(c)), respectively. We note that the sample with x = 0.08 is close to superconducting dome in the phase diagram. 
(b) High-resolution low energy spectra. Here, $k_1$ (on the outer Fermi surface sheet) and $k_2$ (on the inner Fermi surface sheet) denote the Fermi momenta of the two hole bands near $\Gamma$. Insets are the constant energy contours at the Fermi energy.
(c) Corresponding energy distribution curves (EDCs) of the spectra shown in (b). Red dashed lines denote the dip positions.
}
\label{Fig1}
\end{figure}

Three doped samples of FeSe$_{1-x}$S$_x$ with x = 0, 0.04, and 0.08 were measured, and within 100 meV, four bands are clearly resolved near the Brillouin-zone center, taking the FeSe as an example (right panel in Fig.~\ref{Fig1}(a)). The two bands below the Fermi energy are $d_{yz/xz}$ ($\sim-$20 meV) and $d_{xy}$ ($\sim-$50 meV) orbits derived bands, consistent with the previous reports \cite{Watson2015,Gerber2017}. It is generally believed that there is only one hole band of $d_{xz/yz}$ orbit in the Brillouin-zone center in FeSe. However, with improved energy and momentum resolution, two hole bands are identified crossing the Fermi energy, consistent with a recent report in FeSe \cite{Li2020a}, with an estimated splitting energy of about 3 meV. Such two-hole band is evidenced in all the S-substituted samples we measured.
It could be understood by the splitting of the $d_{xz/yz}$ band due to an additional symmetry breaking such as a ferromagnetic order or the Rashba spin-orbit coupling, but still lack of straightforward evidence \cite{Li2020a}.
The anisotropy of the electronic structure is reduced with more S substitution (the constant energy contours in the insets of Fig.~\ref{Fig1}(b)), and interestingly, we find that only the change of the Fermi momentum component along the short axis ($k_x$) of the elliptical Fermi surface sheets is resolvable, suggesting that the nematic phase only affects the electronic structure in one direction (the less conductive direction).

The electronic states near the Fermi energy within 15 meV (photoemission spectra in Fig.~\ref{Fig1}(b) and the corresponding EDCs in Fig.~\ref{Fig1}(c)) clearly show the superconducting coherence peaks at the Fermi crossing of $k_1$ and $k_2$. With a high energy resolution of 0.4 meV, the narrowest FWHM of the peak is about 1.5 meV, which is close to the physical limit for the experimental temperature 4.5 K. In addition, clear peak-dip-hump features are resolved, and the superconducting peak of the $k_1$ extends such a large momentum region to $k_2$ indicates that the observed peak-dip-hump feature cannot be simply due to a result of combination of the superconducting peak and the dispersion of the inner band. A strong electron-boson coupling possibly explains the observation. By subtracting the energy gap from the dip energy, the estimated boson energy is around 1 meV. Although it is hard to determine the origin of this collective mode at such low energy from the current experiment, there is a signature of the dip energy moving to lower binding energy with increasing the S substitution, suggesting the reduction of the energy gap as well \cite{Sandvik2004}.

\begin{figure}
\centering\includegraphics[width=1\columnwidth]{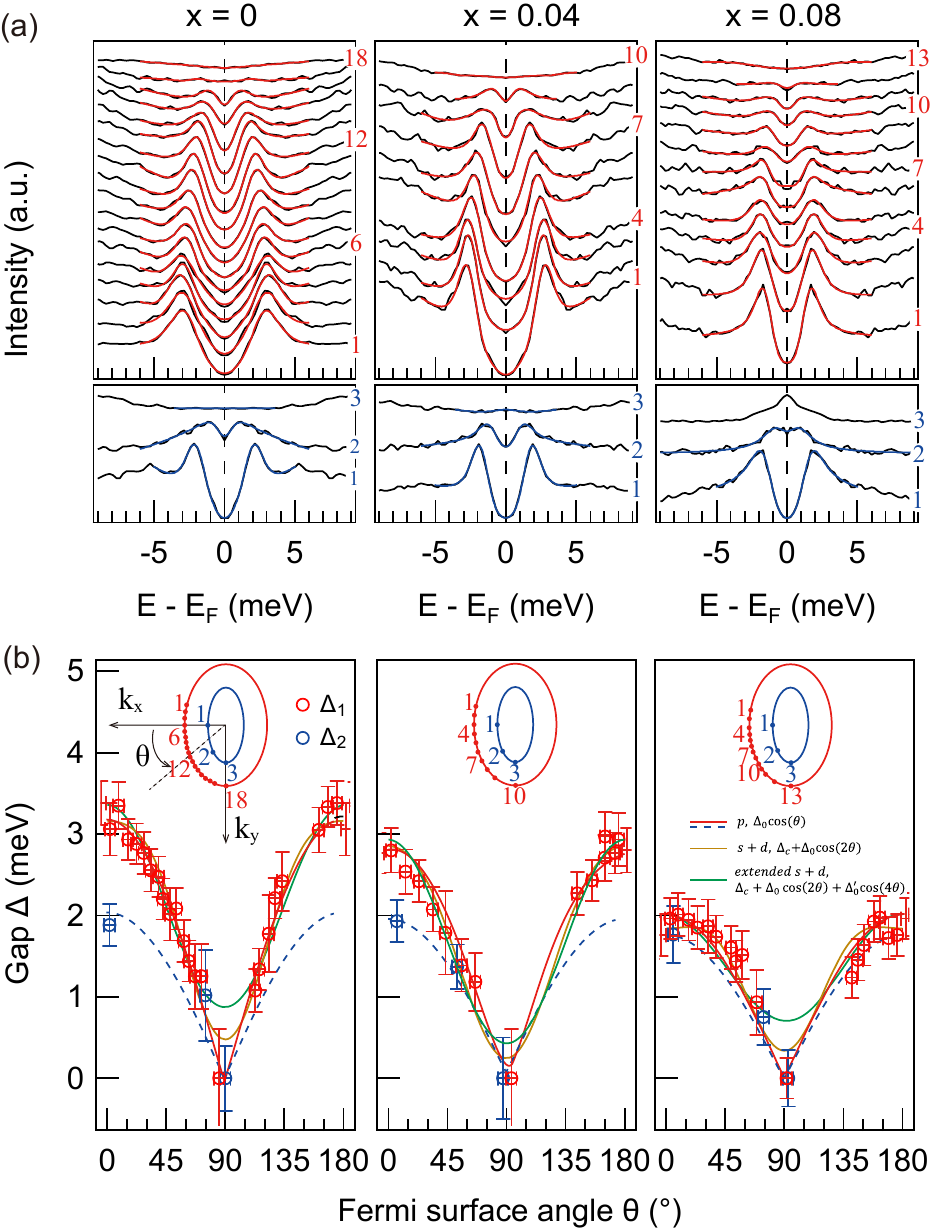}
\caption{
Superconducting gap on the two hole bands of the FeSe$_{1-x}$S$_x$  at 4.5 K.
(a) Symmetrized EDCs at the Fermi momenta along the outer (upper panels) and inner (lower panels) Fermi surface sheets. All curves are fitted by a phenomenological gap function (red and blue curves).
(b) Momentum dependent energy gaps of outer Fermi surface sheet ($\Delta_1$) and inner Fermi surface sheet ($\Delta_2$) extracted from the EDCs shown in (a). The energy gaps are fitted to three twofold-symmetrized pairing gap function: s + d, extended s + d, and p-wave gap functions.
}
\label{Fig2}
\end{figure}

High quality data with ultrahigh energy resolution enables us to track the superconducting gap symmetries  for the two hole Fermi surface sheets with a high precision (Fig.~\ref{Fig2}). The symmetrized EDCs clearly show strongly anisotropic energy gaps for all the dopings of the sample (Fig.~\ref{Fig2}(a)). Firstly, the maximum energy gap of the outer hole is about 3 meV, which is about 50\% larger than the 2 meV gap of the inner hole in the $x$ = 0 sample. Considering the two Fermi surface sheets and two superconducting gaps with different sizes observed here, the origins of the multiple superconducting peaks in the tunneling experiments need to be rechecked \cite{Song2011,Watashige2015,SprauP.2017}. In addition, within our energy resolution, no resolvable energy gap is evidenced at the Fermi momenta on the long axis of the elliptical Fermi surface sheets, indicating node gaps for the two hole bands within the energy resolution (Fig.~\ref{Fig2}(b)). All the energy gaps are two-fold symmetry with resolution limited nodes, consistent with previous reports \cite{Song2011,Liu2018,Rhodes2018,Hashimoto2018}, and can be captured by both of the p and s + d wave gap function within error bars. 
Such anisotropic energy gaps are possibly attributed to the orbital origin and could be a result of the anisotropic electronic states of the nematic phase \cite{Liu2018,Rhodes2018}. 
Experimental energy gap symmetry on sample of more S substitution with no nematic phase will be important in determining the origin of the anisotropic energy gap and also the pairing symmetry.
Moreover, it is clear that the energy gap drops as increasing the x for the outer Fermi surface sheet but keeps changeless for the inner one, indicating an anomalous superconducting gap evolution with S substitution. Although the residue of the spectral intensity from the other domain may bring difficulties in determining the size of the energy gap, the measured sharp superconducting coherence peaks and the apparent coherence peaks between the $k_1$ and $k_2$ ensure the above experimental observation of energy gap evolution.

\begin{figure}
\centering\includegraphics[width=1\columnwidth]{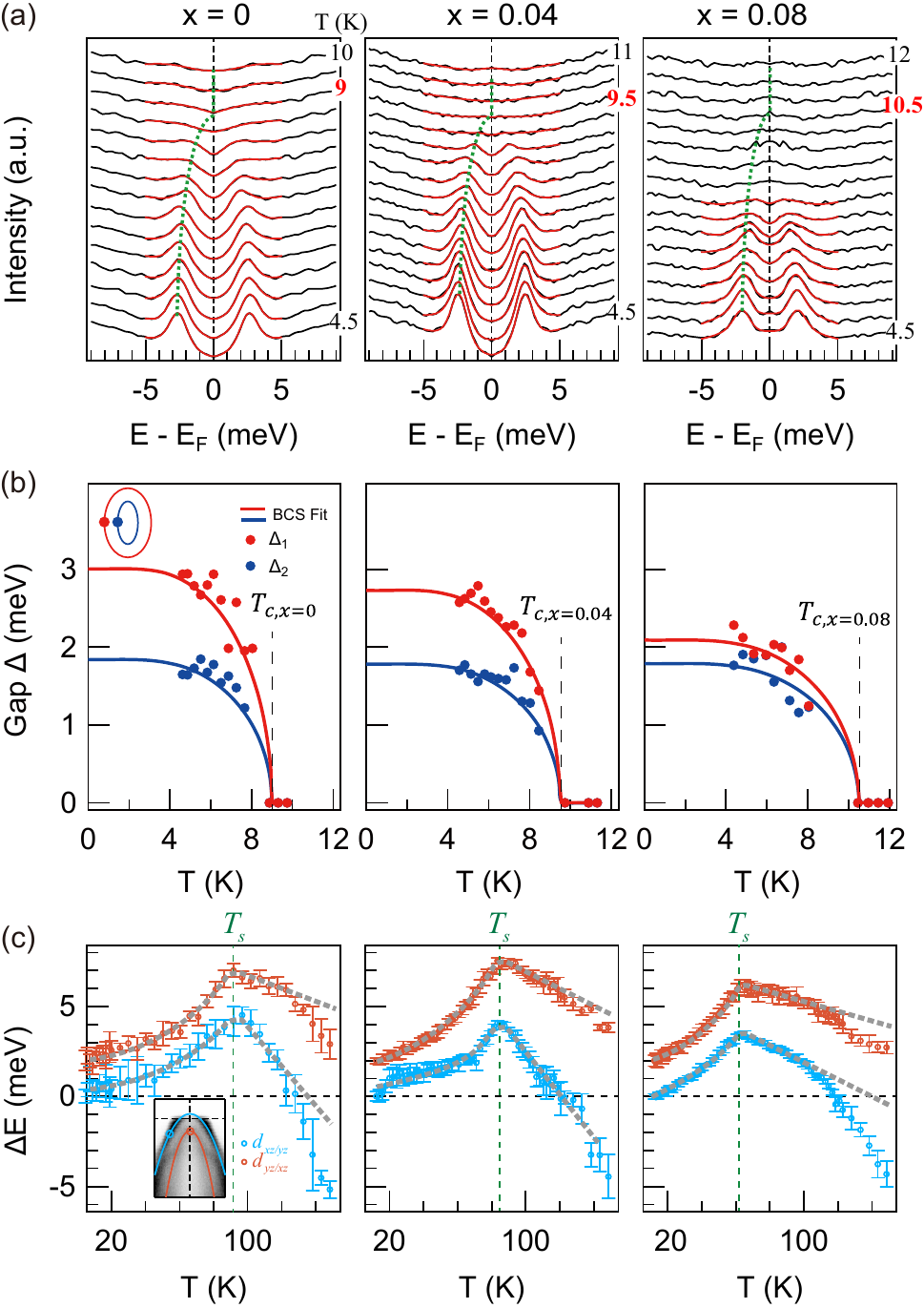}
\caption{
Temperature dependent energy gaps and band structural evolution in FeSe$_{1-x}$S$_x$.
(a) Symmetrized EDCs at the gap maximum point ($\theta$ = 0) of the outer Fermi surface sheet, fitted to a phenomenological gap function (red curves) for x  =0, 0.04, and 0.08.
(b) Temperature dependent superconducting gap of both the outer ($\Delta_1$) and inner ($\Delta_2$) Fermi surface sheets at $\theta$=0. Both the $\Delta_1$ and $\Delta_2$ follow the BCS gap function very well.
(c) Temperature-dependent band shifts of both the $d_{yz/xz}$ and $d_{xz/yz}$ bands for the three different S dopings of the sample.
Grey thick dashed lines are guides to the eyes.
}
\label{Fig3}
\end{figure}

\begin{figure}
\centering\includegraphics[width=1\columnwidth]{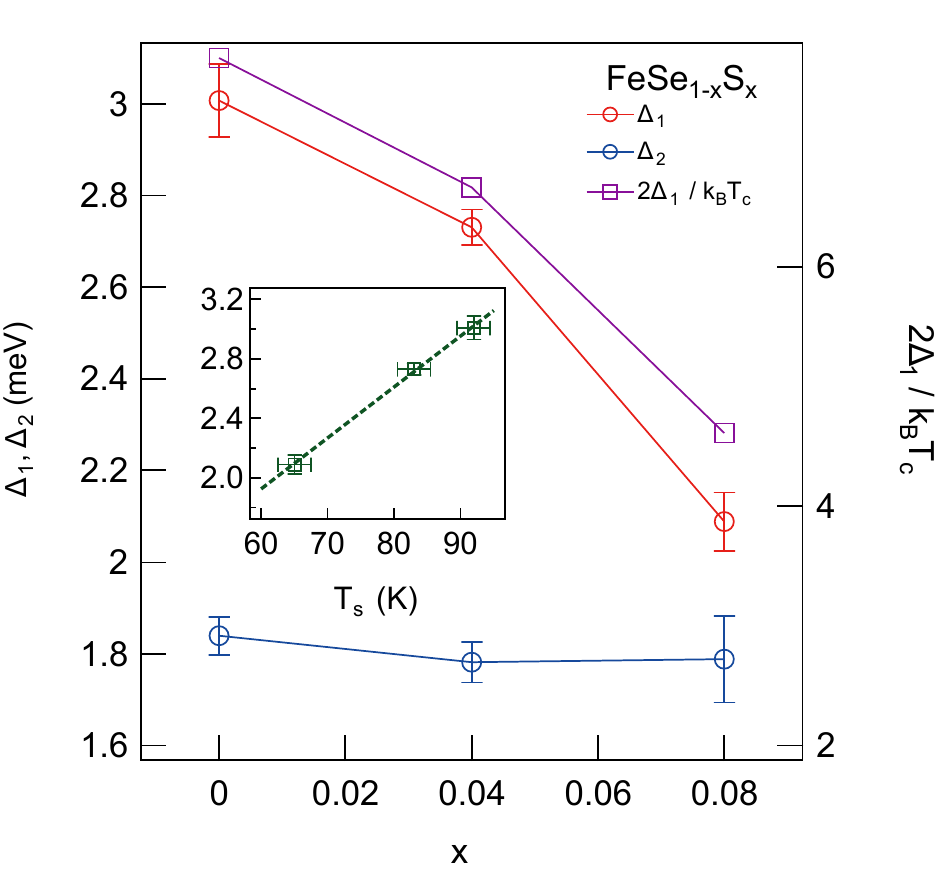}
\caption{
S-substitution dependent energy gap and the $2\Delta_{1}/k_BT_c$ in FeSe$_{1-x}$S$_x$. Inset is the our Fermi surface gap ($\Delta_1$) as a function of nematic or structural transition temperature $T_s$.
}
\label{Fig4}
\end{figure}

The energy gap at zero temperature and the structural or nematic transition temperature ($T_s$) are extracted from the temperature dependent data (Fig.~\ref{Fig3}). The $\Delta_1$ and $\Delta_2$ at zero temperature (in Fig.~\ref{Fig4}) for all the samples are extracted by fitting the temperature dependent energy gaps shown in Fig.~\ref{Fig3}(b) to the BCS self-consistent integral of the superconducting gap $\eta=\int_{0}^{\varepsilon_{0}}\text{tanh}\frac{\sqrt{\Delta^2+\xi^2}}{2k_BT}\frac{d\xi}{\sqrt{\Delta^2+\xi^2}}$ with the giving $T_c$ from transport measurements. All the fittings show that the energy gap as a function of temperature follows the BCS behavior very well, suggesting strong interband coupling between the two holes \cite{Kogan2009}. In addition, there is no resolvable pseudogap in the Brillouin-zone center hole bands, suggesting the claimed pseudogap behavior is possibly in the electronic state at the Brillouin-zone corner \cite{Kang2020}. Clear transitions at the structural or nematic critical temperature ($T_s$) are identified with heating the sample further up to 150 K (Fig.~\ref{Fig3}(c)), giving $T_{s,x=0}$ = 92 K, $T_{s,x=0.04}$ = 83 K, and $T_{s,x=0.08}$ = 65 K with a precision within 5 K. The obtained $T_s$ and $T_c$ match the FeSe$_{1-x}$S$_x$ phase diagram from the previous report \cite{Coldea2019}, suggesting the high quality and precise S substitution of the samples.

Consistent with the momentum dependent energy gap shown in Fig.~\ref{Fig2}(b), the superconducting gap at zero temperature drops by about 50\% for the outer hole pocket from about 3 meV but remains a constant at about 1.8 meV for the inner hole pocket from S-substitution $x$ = 0 to $x$ = 0.08 (Fig.~\ref{Fig4}). However, in contrast, the superconducting transition temperature is increased from 9 K ($x$ = 0) to 10.5 K ($x$ = 0.08). The calculated $2\Delta_\text{1,max}/k_BT_c$ drops from about 7.7 to about 4.6, deviated from the claimed universal value of $2\Delta_\text{max}/k_BT_c$ = 7.2 in various iron-based superconductors  \cite{Miao2018} for higher S substitution.
Considering the BCS theory $\Delta=\epsilon_De^{-\frac{1}{V\cdot\rho_F}}$, in which $\epsilon_D$ is the Debye cut-off energy, V is the attractive potential, and $\rho_F$ is the density of states at the Fermi energy, the reduced $2\Delta/k_BT_c$ suggests weaker coupling (smaller V) with higher S substitution in FeSe$_{1-x}$S$_x$, since the carrier density is increased in the Brillouin-zone center (larger hole pocket with S substitution, insets of Fig.~\ref{Fig1}(b)) as it is also suggested by the quantum oscillation measurement \cite{Coldea2019}. 
In contrast, the $2\Delta_\text{2,max}/k_BT_c$ is close to 4 and nearly S-substitution independent, suggesting unusual interplay between the two holes. Thus, the negative correlation between the superconducting gap and transition temperature is possibly due to smaller density of charge carriers and stronger coupling strength with less S substitution. Further experimental and theoretical studies are necessary to clear this interesting issue.

There may be exotic explanation about the energy gap suppression with S substitution in FeSe, since it seems that the energy gap is positively correlated with the nematic phase. There are a number of proposals that the orbit-dependent spin fluctuations coupled with electrons may be the pairing mechanism in FeSe \cite{Kreisel2017,Benfatto2018,Kreisel2018}, the orbital selectivity can be enhanced by the nematic order \cite{Yu2018}, and recent experiment suggests that the nematicity is likely spin-orbital intertwined \cite{Li2020} and the strongly anisotropic superconducting gap is possibly orbital origin \cite{Rhodes2018}. From the above interplay among superconducting gap, nematicity, orbit, and spin fluctuations, we can conclude that strong electronic nematicity favors to enhance the pairing strength, as shown in the inset of Fig.~\ref{Fig4} that the size of the superconducting gap is proportional to the nematic transition temperature, although the superconducting transition temperature does not scale to the superconducting gap necessarily.
This scenario is consistent with the superconducting gap in FeSe$_{1-x}$Te$_x$ with x = 0.55, at which doping the nematic transition temperature is nearly 0, so that the measured superconducting gap is even smaller although the $T_c$ is higher \cite{Zhang2018}. Our observation is also qualitatively consistent with the much smaller energy gap from the transport measurement in FeS, in which there is no nematic order \cite{Xing2016}.

Now we turn back to the physical origin of the two holes observed at low temperature in FeSe$_{1-x}$S$_x$. For a Rashba spin-orbit band splitting scenario, the theoretically estimated superconducting gap variation of two spin-polarized bands is less than 5\% \cite{Weng2016}. However, the observed energy gap difference between the two hole bands is about 50\%, which is one order larger. In addition, there is no photoemission circular dichroism observed by tuning the polarization of the probe beam in our measurement (data not shown), strongly suggesting that the two-hole bands cannot be a result of Rashba splitting. A ferromagnetic order, which breaks the time-reversal symmetry, may induce such band splitting around 3 meV, but no ferromagnetic order has been identified in FeSe$_{1-x}$S$_x$ so far \cite{Wang2016}. Spin polarized measurement with high energy resolution and theoretical study are required to clear this issue.

In summary, we have discovered unusual two-hole Fermi surface sheets near the Brillouin-zone center in FeSe$_{1-x}$S$_x$ with x up to 0.08, and resolution limited two-fold highly anisotropic gaps on the two hole pockets. The reduction of the energy gap in the outer hole pocket with S-substitution increment suggests that the appearance of nematicity enhances the pairing strength possibly due to orbital dependent spin fluctuation. However, the superconducting transition temperature does not scale with the size of the energy gap. The above observations strongly constrains the theory on the electron pairing mechanism and also the origin of the observed two hole pockets in FeSe$_{1-x}$S$_x$, and study with heavier S-substitution is necessary to track the unusual band structure and also the superconducting gap symmetry further. 

\begin{acknowledgments}
W.T.Z. acknowledges support from the Ministry of Science and Technology of China (Grant No. 2021YFA1401800) and National Natural Science Foundation of China (Grant No. 11974243 and 12141404) and additional support from a Shanghai talent program. D.Q. acknowledges support from the National Natural Science Foundation of China (Grant No. 12074248) 
The work at Fudan University was supported by the Innovation Program of Shanghai Municipal Education Commission (Grant No. 2017-01-07-00-07-E00018) and the National Natural Science Foundation of China (Grant No. 11874119)

\end{acknowledgments}

\bibliographystyle{apsrev4-2}

\end{document}